# Evaluating the Magnetorotational Instability's Dependence on Numerical Algorithms and Resolution

By

Dinshaw S. Balsara[1] (dbalsara@nd.edu) and Chad D. Meyer[1] (cmeyer8@nd.edu)

**Key Words:** turbulence, MHD, methods: numerical, accretion discs, instabilities

**Mailing Address:**

1:  Department of Physics
    225 Nieuwland Science Hall
    University of Notre Dame
    Notre Dame, Indiana 46556, USA

**Phone:** (571) 631-2712
**Fax:** (571) 631-5942

**Abstract**

Most quantitative results associated with the magnetorotational instability (MRI) have been obtained exclusively via numerical experiments. In a recent paper Fromang & Papaloizou (2007) used an isothermal version of the ZEUS code to show that the saturated results from MRI simulations depend strongly on the details of the numerical algorithm and also on the resolution of the simulations. The code they employed relies on a staggered mesh, non-conservative algorithm with an artificial viscosity formulation. Higher order Godunov schemes for magnetohydrodynamics (MHD) have come of age. These schemes offer conservative, entropy-satisfying formulations along with a range of reconstruction strategies. The small-scale dissipation is also better controlled with Riemann solvers, and several such Riemann solvers have become available. The RIEMANN code implements a number of these algorithms. For this reason, we examine the dependence of the saturated MRI-driven turbulence on several higher order Godunov algorithms and also on the numerical resolution used for simulating it.

Since second order accurate Godunov algorithms seem to be the current mainstay in astrophysics research, we restrict our attention to these schemes. We study the role of MinMod, Monotone Centered (MC) and Weighted Essentially Non-Oscillatory (WENO) reconstruction strategies. The weight of experience suggests that MinMod is the most dissipative reconstruction strategy while WENO is the least dissipative. Even amongst the second order accurate Godunov schemes, one can use a range of Riemann solvers, the popular choices being the HLLE, HLLD and linearized Riemann solvers. The HLLE Riemann solver detects fast magnetosonic waves while averaging over the other wave families that are present in the MHD hyperbolic system; the HLLD Riemann solver additionally recognizes the presence of Alfven waves; the linearized Riemann solver further retains the slow magnetosonic waves. By carrying out simulations on a range of resolutions, we show that the details of the isothermal MRI-driven turbulence depend principally on the Riemann solver and secondarily on the reconstruction strategy. Use of the HLLE Riemann solver consistently causes the isothermal MRI-driven turbulence to die out. By using HLLD or linearized Riemann solvers, the simulations can sustain a robust MRI-driven turbulence if the reconstruction strategy is adequately sophisticated. The wave structure that is built into the Riemann solver is, therefore, clearly shown to control the effective magnetic Prandtl number and, therefore, influence outcomes.

We have studied saturated, MRI-driven turbulence using three-dimensional, isothermal simulations with resolutions that extend from $64^3$ zones to $192^3$ zones. We find that the effective $\alpha$ parameter tends to show progressively smaller decrements with increasing resolution when the best reconstruction strategy (WENO) and the best Riemann solver (linearized) are used. We attribute this result to the more sophisticated dissipation mechanisms that are used in higher-order Godunov schemes. Spectral analysis and transfer functions have been used to quantify the dissipative processes in these higher-order Godunov schemes.

**I) Introduction**

Ever since the classic paper of Shakura & Sunyaev (1973), there has been a great deal of interest in finding effective mechanisms for angular momentum transport in accretion disks. The above authors conjectured that the angular momentum transport takes place as a result of a viscosity in the accretion disk's fluid. To sustain observed rates of accretion, they had to rely on a viscosity parameter that was dramatically larger than the viscosity that could be provided by molecular effects. It was, therefore, conjectured that the effective viscosity is provided by a macroscopic turbulence in the accretion disk. Thus Shakura & Sunyaev (1973) had to posit an effective scale-free $\alpha$ parameter so that $\alpha\, c_s\, H$ serves as the coefficient of the viscous terms in an accretion disk. Here $c_s$ is the local sound speed and H is the local scale height of the accretion disk. Values of $\alpha$ that lie in the range of 0.001 to 0.01 are needed in order to explain the measured, quiescent accretion rates on to T Tauri stars and various compact objects (Hartmann 2008, Warner 1995, Lewin, van Paradijs and van den Heuvel 1995).

While fluid dynamic processes did not yield a mechanism for persistently driving the turbulence in accretion disks, magnetohydrodynamical (MHD) processes stood a better chance (Velikhov 1959, Chandrasekhar 1960, Balbus & Hawley 1991, 1998). In a real accretion disk, the matter is a plasma and is permeated by magnetic fields. Stretching magnetic fields can cause them to have a tensional force. The magnetic fields connect lumps of matter on adjacent Keplerian orbits. As the lumps of matter drift apart due to differential rotation, their distance increases. This results in the magnetic fields being stretched. The resulting tensional force causes angular momentum transport between the adjacent lumps. If this process of stretching the fields is repeated often enough, substantial amounts of angular momentum transfer will take place, resulting in efficient accretion of matter on to the central star. This paradigm is known as the magneto-rotational instability (MRI) and had been developed in the literature over the years. It is interesting to note that this field is largely been based on numerical simulations (Hawley & Balbus 1991, Hawley *et al.* 1995, Brandenburg *et al.* 1995, Miller & Stone 2000, Steinacker & Papaloizou 2002, Fromang & Nelson 2006). Such simulations tend to be compressible and transonic, reaching local Mach numbers as high as about 0.6. As a result, it is only natural to simulate them with shock-capturing schemes. Such simulations also tend to be very long-running because the actual, stable MRI-driven turbulence only emerges after the initial transient fluctuations in the flow have died out. Depending on the details of the numerical scheme used and the initial conditions that were supplied to it, this damping of initial transients may take as much as thirty to forty orbits. Since the dissipation of turbulent energy in an adiabatic simulation would generate a large amount of heating in the gas, such simulations have usually been done with isothermal or polytropic gas laws. Goodman & Xu (1994), Latter, Lesaffre & Balbus (2009) and Lesaffre, Balbus & Latter (2009) have also studied the role of various channel modes that tend to develop in MRI simulations using analytical and numerical approaches. In recent years there has also been a trend to use the MRI as a tool for studying other disk-related processes. Thus the role of two fluid MHD has been studied (Hawley & Stone 1998) and the importance of ionization in protostellar disks has been studied recently (Turner, Carballido & Sano 2009). Radiative effects in disks have been studied in Flaig, Kissman & Kley (2009). MRI simulations have also been used to study dust sedimentation in protostellar accretion disks (Carballido *et al.* 2006, Johansen *et al.* 2007, Balsara *et al.* 2009a, Tilley *et al.* 2010).

Fromang & Papaloizou (2007) (FP07 henceforth) published a very important paper where they carried out isothermal MHD simulations of MRI-driven turbulence on a sequence of meshes with increasing resolution. The simulations were carried out for a large number of orbits in a local shearing box approximation (Goldreich & Lynden-Bell 1965, Hawley *et al.* 1995). While the turbulence saturates in thirty to forty orbits, FP07 ran the ZEUS code on this problem for almost a hundred orbits. This allowed them to ascertain the saturated state of the MRI turbulence. FP07 found that the fully-developed turbulence arising in MRI simulations depends strongly on the details of the numerical algorithm used and also on the resolution of the simulations. In particular, they found that the effective $\alpha$ drops by a factor of two with each doubling of the resolution in their simulations. This resolution-dependence, coming as it did after years of research on the MRI with the same set of numerics, was found to be a very perplexing result. FP07 used this to conclude that the numerical dissipation in their code was inadequate when it came to producing resolution-independent results. Falle (2002) has carried out an analysis of the dissipation characteristics of that code. Clarke (1996) has reported that such numerical algorithms are prone to "explosive" growth (within a single time step) of weak magnetic fields to possibly dynamic strengths in the vicinity of strong local shear. To keep this paper forward-looking, we avoid further discussion of the numerics in FP07 as much as possible. FP07 concluded that physical dissipation could resolve this difficulty. Umurhan et al. (2007a,b) have presented an elegant analytic theory suggesting that the effective $\alpha$ for MRI-driven turbulence could be dependent on the magnetic Prandtl number. The magnetic Prandtl number is defined as $\mathrm{Pm} \equiv \nu/\eta$ where $\nu$ is the kinematic viscosity and $\eta$ is the resistivity. Fromang *et al.* (2007, FPLH07 henceforth) and Lesur & Longaretti (2007) used numerical simulations to show that for each choice of the Reynolds number there exists a critical Pm below which the MRI-driven turbulence dies. Here the Reynolds number $\mathrm{Re} \equiv c_s H/\nu$ where $c_s$ is the local sound speed and H is the local scale height of the gas in the disk. FPLH07 found that the critical Pm is a little more than unity for a large range of Re.

One could, in fact, include a physical viscosity and resistivity in all MRI simulations. Doing so, however, carries certain consequences. First, there is the conceptual difficulty associated with choosing one specific magnetic Prandtl number over another. A consideration of molecular viscosities and resistivities in accretion disks would show that they have an extremely small value of Pm except within a few Schwarzschild radii of a compact object (Balbus & Henri 2008). If molecular processes were to be used, the resulting value for the effective $\alpha$ would not even be close to the range that is required for accretion processes in TTauri stars and compact objects. Since this paper has a numerical focus, we do not address this issue any further here. Second, there is the numerical difficulty associated with raising the physical viscosities and resistivities to levels where they are clearly larger than the native numerical viscosity and resistivity of the shock-capturing scheme. Toro (1992) has shown that when a physical viscosity and a numerical viscosity are simultaneously included in a numerical code, the physical viscosity should always exceed a certain amount that is proportional to the local sound speed times the zone size. If such a condition is not met, the numerical viscosity will overwhelm the physical viscosity. Toro (1992) constructs limiters that include the role of a physical viscosity. However, since that work was not developed any further, it is not quite ready for use in production codes. Consequently, dialing in a pre-specified viscosity and resistivity into a shock-capturing numerical code often results in a substantial loss of resolution. When the focus is solely on a theoretical study of the MRI, this may still be an acceptable state of affairs. For applications where one wants to use the MRI to explore other disk-related physics, this loss of resolution might be unacceptable. It is, therefore, useful to understand the performance of different types of numerical algorithms when they are applied to the MRI. The full gamut of numerical schemes that are used in computational astrophysics is indeed very large. We undertake such a study here within the context of higher order Godunov schemes.

Higher order Godunov schemes for numerical MHD have now reached maturity. Most such algorithms have been restricted to second order of accuracy (Dai & Woodward 1994, Powell 1994, Ryu & Jones 1996, Balsara 1998a,b, Falle, Kommissarov & Joarder 1998, Daedner *et al.* 2002, Crockett *et al.* 2005, Ustyugova *et al.* 2009) though some have sought to go beyond second order accuracy (Dumbser *et al.* 2008, Li 2008, Balsara *et al.* 2009b, Balsara 2009a). The magnetic fields in such algorithms are evolved according to Faraday's law, the structure of which is such that it ensures that once the magnetic field is divergence-free, it remains so for all time. The utility of schemes that preserve this divergence-free constraint in the magnetic field has been discussed by several authors (Yee 1966, Brecht *et al.* 1981, Evans & Hawley 1989, DeVore 1991). Because of the periodicity and long running times in local MRI calculations, such constraint preserving schemes might be especially suitable for this type of calculation. Several papers have used conventional one-dimensional Riemann solvers for implementing a discrete version of Faraday's law in higher order Godunov schemes (Ryu *et al.* 1998, Balsara & Spicer 1999, Balsara 2004, Londrillo & DelZanna 2004, Gardiner & Stone 2005). Recently, Balsara (2010) has formulated multidimensional Riemann solvers, which should go a long way in removing the ambiguity in updating the magnetic field. Since conventional, second-order accurate, divergence-free, Godunov schemes for MHD are becoming the mainstay in computational astrophysics, we use them in this paper to analyze the MRI.

While the above-mentioned second order accurate Godunov methods might differ in details, they are all built from two common ingredients – a reconstruction strategy and a Riemann solver. The reconstruction strategy imparts the flow variables within each zone with internal structure. This internal structure may often be a piecewise linear profile, but can on occasion assume a parabolic profile. This profile is limited, i.e. certain non-linear functions are used to ensure that the profile does not develop any new extrema at shocks. This limiting procedure prevents the development of post-shock oscillations. In this work we explore the role of different limiters. The MinMod limiter is the least sophisticated. It simply minimizes the left and right-sided slopes that are evaluated from either side of a zone, setting the slope to zero when left and right-sided slopes carry opposite signs. The Monotone Centered (MC) limiter is a fair bit more sophisticated because it includes the centered slope as one of the choices and permits some latitude in the definition of a slope. The Weighted Essentially Non-Oscillatory (WENO) limiter draws on the centrally biased slopes arising from the r=3 WENO scheme of Jiang & Shu (1996). While the MinMod and MC limiters clip extrema, the WENO scheme does not, which permits smooth profiles to retain their structure. Just as numerical schemes can have different reconstruction strategies, they can also have different Riemann solvers that are built on a variety of wave models. The wave model for the MHD system has been analyzed with a view to numerical usage in Roe & Balsara (1996) for adiabatic MHD and Balsara (1996) for isothermal MHD. The HLLE Riemann solver (Harten, Lax & van Leer 1983, Einfieldt *et al.* 1991, Janhunen 2000) preserves only the fast magnetosonic waves while averaging over the other wave modes. The HLLD Riemann solver (Miyoshi & Kusano 2004, Mignone 2007) utilizes a wave model that exactly represents fast magnetosonic waves and also Alfven waves. The linearized Riemann solver (Roe 1981, Cargo and Gallice 1997, Balsara 1998a) represents all the wave modes that develop in a one-dimensional analysis of the MHD system. Consequently, the HLLE Riemann solver is the least expensive Riemann solver but it is also lacks a sophisticated wave model. The HLLD Riemann solver introduces some additional sophistication in its wave model at the expense of an increased computational cost. The linearized Riemann solver for MHD carries a slightly higher computational cost than the HLLD Riemann solver but provides a very refined wave model in return. There are minor differences in methods that are used for achieving second order temporal accuracy but, if everything is done right, they are all equivalent. Similarly, higher order Godunov methods use different strategies for stabilizing very strong shocks but, owing to the low Mach numbers in the MRI, the differences are immaterial. We hope, therefore, that this paragraph provides all readers with adequate detail to understand the different algorithms that are intercompared. All of these algorithms have been implemented in the RIEMANN code, which is being readied for public release with the first author's text on computational astrophysics.

With these methods in hand, we run the same isothermal, cubical simulation on a sequence of meshes ranging from $64^3$ to $192^3$ zones. The simulations have been run for a hundred orbits so that the initial transients have certainly died out and a time-steady MRI, if it chooses to develop, has indeed developed. Different combinations of a reconstruction strategy and a Riemann solver are tried. Scalar and spectral diagnostics are used to analyze the results. Section II describes the simulations. Section III presents scalar diagnostics for all our runs. Section IV catalogues the time-variation of the MRI. Section V analyzes the dissipation characteristics of the code using spectral diagnostics. Section VI offers a discussion and some conclusions.

**II) Description of the Simulations**

We perform a series of simulations in unstratified shearing-box geometry to model the MRI driven turbulence and evolve the gas using the equations of ideal, isothermal MHD. The *x*, *y*, and *z* directions in the simulations correspond to the radial, azimuthal and vertical directions respectively. In this geometry, the governing equations can be written as

$$\frac{\partial \rho}{\partial t} + \nabla \cdot (\rho \mathbf{v}) = 0 \qquad (1)$$

$$\frac{\partial \mathbf{v}}{\partial t} + (\mathbf{v} \cdot \nabla)\mathbf{v} + 2\mathbf{\Omega} \times \mathbf{v} = -\frac{1}{\rho}\nabla\left(\rho c_s^2 + \frac{\mathbf{B}^2}{8\pi}\right) + \frac{(\mathbf{B}\cdot\nabla)\mathbf{B}}{4\pi\rho} + 2q\Omega^2 x \hat{\mathbf{x}} \qquad (2)$$

$$\frac{\partial \mathbf{B}}{\partial t} = \nabla \times (\mathbf{v} \times \mathbf{B}) \qquad (3)$$

where $\rho$ is the gas density, **v** is the velocity, and **B** is the magnetic field.

We simulate a cubic domain representing a physical patch of the accretion disk extending over $[-H/2, H/2]^3$, where $H = \sqrt{2}c_s/\Omega$ is the scale height of the gas. The boundaries are periodic in the *y* and *z* directions and periodic in the shearing coordinates in the *x* direction (Goldreich & Lynden-Bell 1965, Hawley *et al.* 1995).

For all of our runs, we have fixed the isothermal sound speed $c_s = \sqrt{0.5 \times 10^{-6}}$, $\Omega = 1.0 \times 10^{-3}$, and initial density $\rho_0 = 1.0$. The velocities are initially perturbed from the purely background shear by randomly phased fluctuations with amplitudes 5 per cent of the isothermal sound speed. The magnetic fields are initialized by a vector potential

$$A_x = \left(32\rho_0 c_s^2 / \pi\beta_0\right)^{1/2} H\cos(2\pi y/H)\cos(2\pi z/H) \qquad (4)$$

where $\mathbf{B} = \nabla \times \mathbf{A}$. In light of eqn. (4), our simulations have no net magnetic flux. We have chosen the initial ratio of thermal to magnetic pressure $\beta_0 = 40$. In cases where a persistent channel mode developed, we re-ran the simulations with $\beta_0 = 400$. In all such cases we found that the channel mode formed regardless of the value of $\beta_0$. All simulations were run for a hundred orbits.

Table I introduces the nomenclature of the runs as well as their resolution. The linearized, HLLD and HLLE Riemann solvers are labeled by "L", "D" and "E" respectively and form the first letter in the names of the runs. The WENO, MC and MinMod reconstruction strategies are denoted by "W", "C" and "M" respectively and constitute the second letter in the name of each run. The resolution in any one direction yields the third tag in the name of the run. Thus "LW128" denotes a run with linearized Riemann solver and WENO reconstruction at a resolution of $128^3$ zones.

### III) Analysis of Scalar Variables

The MRI-driven turbulence produces stresses that drive mass inwards toward the star while driving angular momentum away from the star. As a result, it is not isotropic turbulence but rather a turbulence that produces stresses with a preferential sign. In keeping with Hawley et al. (1995), we define an effective $\alpha$-coefficient for the Reynolds stresses, the Maxwell stresses and the combination of the two as follows

$$\alpha_{\text{Rey}} = \frac{1}{P_0} \left\langle \rho (v_x - \overline{v}_x)(v_y - \overline{v}_y) \right\rangle \quad (5)$$

$$\alpha_{\text{Max}} = \frac{1}{P_0} \left\langle -\frac{B_x B_y}{4\pi} \right\rangle \quad (6)$$

$$\alpha_{\text{Tot}} = \alpha_{\text{Rey}} + \alpha_{\text{Max}} \quad (7)$$

Here $\overline{v}_x$ and $\overline{v}_y$ are the local values of the mean flow. It is also useful to evaluate the volume-averaged, saturated kinetic and magnetic energies. In all our simulations we found that the MRI-driven turbulence establishes itself in about forty orbits. We will justify this observation in the next section. Time-averaged values for all of the above-mentioned scalar values are presented in Table II, where the time-averaging extends from the fortieth orbit to the end of the simulation.

Let us first focus on the simulations that have been labeled LW, i.e. they were carried out with a linearized Riemann solver and the WENO limiter. We see from Table II that $\alpha_{Max}$ always exceeds $\alpha_{Rey}$ with the result that most of the effective viscosity derives from magnetic stresses. This result is consistent with several previous simulations including those in FP07. Additionally, FP07 found that $\alpha_{Tot}$ decreased with increasing resolution. Their $\alpha_{Tot}$ was inversely proportional to the number of zones per scale height. We see that there is a small reduction in $\alpha_{Tot}$ when going from LW64 and LW96 to LW128. However, the decrement in $\alpha_{Tot}$ becomes even smaller when going from LW128 to LW192; $\alpha_{Tot}$ does not decrease by a factor of 1.5 when going from a $128^3$ zone simulation to a $192^3$ zone simulation. Notice too that LW192 has three times as much resolution in any one direction as LW64, however, $\alpha_{Tot}$ does not decrease by a factor of three for the larger simulation. The slightly higher values of $\alpha_{Tot}$ in the lower resolution simulations can be explained by the fact that all second order accurate Godunov codes rapidly dissipate Alfven waves with wavelengths of fifteen or thirty times the zone size. The extent of the dissipation depends on the numerical algorithm (Balsara 2004). As the order of accuracy of Godunov schemes is increased, the ability to retain small wavelength structures is much-improved (Balsara *et al.* 2009). MRI-driven turbulence that is represented on a $64^3$ zone mesh and simulated with a second order accurate scheme would not develop an inertial range. Many of the magnetic features on such a mesh should be regarded as small-scale structures that are smaller than the dissipation scale in the simulation. They are, therefore, rapidly damped in such simulations. Consequently, only the larger and more coherent structures survive in the lower resolution simulations and they produce larger values of $\alpha_{Tot}$. Notice, though, that if a well-developed turbulence with a large inertial range forms, then $\alpha_{Tot}$ is likely to become a scale-free parameter of the turbulence for each particular choice of Pm. This is equivalent to the expectation from FPLH07 that at large values of Re the critical value of Pm above which a self-sustaining MRI can develop should asymptotically become a constant; see their Figs. 11 and 13. In a higher order Godunov scheme the limiters and Riemann solvers conspire to provide strong dissipation to features whose wavelength approaches a zone width while trying to leave all other features mostly intact. Therefore, as the resolution is increased, we expect the effective value of Re to increase to the point where it assumes rather large values in our simulations. The fact that LW128 and LW192 both yield roughly similar values of $\alpha_{Tot}$ is consistent with the formation of a well-developed, MRI-driven turbulence at these higher resolutions. Observe too that those two simulations produce quite comparable values for the time-averaged magnetic and kinetic energies. It is also interesting to observe that the ratio of time-averaged turbulent magnetic energy to the time-averaged, turbulent, poloidal kinetic energy is comparable to $\alpha_{Max}/\alpha_{Rey}$ in the well-resolved simulations. Thus, the magnetic and kinetic stresses contribute to $\alpha_{Tot}$ in direct proportion to the turbulent energy that they bear.

The above-mentioned trend of having a slightly decreasing $\alpha_{Tot}$ with increasing resolution is also apparent in the LC runs. As before, $\alpha_{Tot}$ does not change by a factor of three when going from LC64 to LC192. We do however see that LC128 and LC192 show a larger change in the value of $\alpha_{Tot}$, a fact which is consistent with the larger dissipation in the MC limiter. The MinMod limiter is known to be substantially more dissipative than the MC and WENO limiters. As a result, the LM64 run does not sustain an MRI and the LM runs do show a dramatic change in $\alpha_{Tot}$ with increasing resolution. Our study of the transfer functions in Section V.2 will further reinforce this conclusion. The MinMod limiter is almost never used for production work in higher order Godunov codes because of its very dissipative treatment of short wavelength modes. However, for the purposes of this work, it plays a very useful role by demonstrating that one can indeed arrive at numerical schemes that show substantial resolution-dependence, as was observed by FP07. Notice too that LW192, LC192 and LM192 all converge to reasonably similar values of $\alpha_{Tot}$. All the limiters permit long wavelength modes to propagate with minimal dissipation. As a result, we see that once the simulation is carried out on a mesh with sufficiently large number of zones, the specifics of the limiter do not strongly influence the eventual value of $\alpha_{\text{Tot}}$. The higher quality limiters do, however, have the virtue of minimizing the damping of waves on intermediate length scales whereas the MinMod limiter produces substantial amounts of dissipation on those scales. The better limiters, therefore, do a substantially better job of obtaining a converged result.

The simulations that were carried out with the HLLD Riemann solver show the same trends as the ones that used the linearized Riemann solver. Mirroring the previous trends, we see from Table II that DW128, DC128 and DM128 seem to converge to almost the same value of $\alpha_{\text{Tot}}$. The HLLD Riemann solver is computationally less expensive than the linearized Riemann solver by a small factor and we see in this paper that its use can also permit the formation of a vigorous MRI-driven turbulence. Notice too that DW128 produces a slightly higher value of $\alpha_{\text{Tot}}$ than LW128. Drawing on the theory of Umurhan et al. (2007a,b), this would imply that the HLLD Riemann solver produces a slightly higher magnetic Prandtl number than the linearized Riemann solver. Strictly speaking, the analysis of Umurhan et al. (2007a) only applies in the limit of very small magnetic Prandtl numbers. However, as mentioned by Umurhan et al. (2007a), it is possible to extrapolate their results to larger magnetic Prandtl numbers. Such an extrapolation gives us a useful perspective on the inner workings of Riemann solvers and how they influence the development of the MRI. The purpose of a Riemann solver is to provide physically-based, small-scale dissipation. A good Riemann solver should strongly damp large amplitude waves and discontinuities whose extent is comparable to the mesh size while leaving flow structures with larger wavelengths undamped. The specific amounts of dissipation that a Riemann solver provides to magnetic fluctuations and velocity fluctuations can differ. The dissipation model used in a Riemann solver determines the relative amounts of dissipation that it provides to the magnetic and velocity fluctuations. Since these dissipation models have a highly nonlinear dependence on the flow structures, the exact amount of dissipation is difficult to quantify. However, it is worth mentioning that the linearized Riemann solver used here included an entropy fix for the Alfven waves, whereas no such fix was applied to the Alfven waves in the HLLD

Riemann solver. Since the entropy fix constitutes an extra source of dissipation, we now understand at a qualitative level why the HLLD Riemann solver produces a slightly larger effective Pm and a correspondingly larger value of $\alpha_{Tot}$. This discussion has also brought out the important insight that the Riemann solver plays a primary role in setting the effective Pm in such simulations.

Our last observation from scanning Table II pertains to the use of the HLLE Riemann solver. This Riemann solver is indeed very computationally efficient. However, it is known to produce substantial amounts of dissipation on a range of length scales. As a result, we see from Table II that channel modes develop very persistently when this Riemann solver is used. Lesaffre, Balbus & Latter (2009) have shown that persistent channel modes develop in cubical domains because smaller parasitic modes do not destroy the channel mode rapidly enough. While we did not see such persistent channel modes in some of our higher quality and higher resolution simulations, we did see them when the HLLE Riemann solver was used. This shows that the HLLE Riemann solver, with its substantially higher dissipation, efficiently damps out the parasitic modes, thus permitting a persistent channel mode to emerge. If we draw on the perspective provided by FPLH07, we can also conclude that the HLLE Riemann solver produces an effective Pm that is less than or equal to unity. Because of the strong non-linearities that are built into higher order Godunov schemes, it is not possible to identify the specific features in the HLLE Riemann solver that cause it to have a smaller magnetic Prandtl number. However, the inability of the isothermal HLLE Riemann solver to sustain an MRI is a strong indicator that it results in an effective Pm $\leq$ 1. The only other situations where we saw the emergence of channel modes were LM64 and DM64, where the combination of low resolution and a MinMod limiter produced a higher level of numerical dissipation which also eliminated the parasitic modes of the MRI.

It is also worth pointing out that our observations about the deficiency of the HLLE Riemann solver only pertain to the isothermal version that was used in this work. In adiabatic calculations it is possible to exchange energy between the magnetic fluctuations and the pressure fluctuations, an energy exchange that is not as efficient in isothermal calculations. As a result, the HLLE Riemann solver might show itself to be quite serviceable for adiabatic simulations of the MRI. Such adiabatic simulations are, however, not being reported on in this paper.

The above paragraphs enable us to conclude that using a higher quality reconstruction, along with a high quality Riemann solver, are essential if one wants to produce a robust MRI turbulence. Furthermore, even with these improvements, the changes in the value of $\alpha_{Tot}$ seem to become very small with increasing resolution only when the number of zones meets or exceeds 192 zones per scale height.

**IV) Time-Variation of the MRI**

Figs. 1a, 1c, 1e and 1g show the variation of the kinetic energy (dotted line), magnetic energy (dashed line) and total energy (solid line) as a function of time in the LW192, LW128, LW96 and LW64 simulations respectively. Figs. 1b, 1d, 1f and 1h show the variation of $\alpha_{Rey}$ (dotted line), $\alpha_{Max}$ (dashed line) and $\alpha_{Tot}$ (solid line) as a function of time for the same set of simulations. We see that these plots are consistent with the time averages and their variances from Table II. In other words, we can visually verify that LW64 and LW96 have higher time-averaged mean values and larger variances in those values than LW128 and LW192. We also see that the initial transients in our effective $\alpha_{Tot}$ have died out within forty orbits, and the same is true for the other scalar variables in Fig. 1. This justifies our decision in the previous Section to evaluate temporal averages from that time onwards. Fig. 1 also visually confirms our claim that $\alpha_{Tot}$ only undergoes modest decreases with increasing resolution. We also see that the runs LW128 and LW192 produce time-averaged values of $\alpha_{Tot}$ that do not differ by 1.5 – the ratio of their linear resolutions.

Observe that the temporal fluctuations in $\alpha_{Tot}$ decrease with increasing resolution. Such fluctuations have been seen in all previous MRI-driven turbulence simulations (e.g. Hawley et al. 1995) and have been interpreted as intermittencies that develop in the turbulence. A low resolution simulation, like LW64 or LW96, can only have a very small number of intense, intermittent events. A larger simulation, like LW128 or LW192, can sustain several such intermittent events that temporally overlap each other, which causes the fluctuations to be smaller.

## V) Spectral Analysis

### V.1) Power Spectra

Fig. 2 displays a snapshot of the x and y velocity and magnetic field magnitudes for LW192 after eighty orbits have been simulated. Each of the panels shows us that an active and vigorous MRI-driven turbulence with a large amount of sub-structure has established itself on all scales in our larger simulations. It is also worth noting that the y-component of the magnetic field has been stretched in the direction of the shear. In the language of turbulence, establishing turbulent fluctuations on a range of scales is tantamount to the formation of a well-defined inertial range which can be distinguished from the dissipation scales. It is reasonable to expect that such an inertial range can be established in an MRI simulation because the source term associated with the centripetal force, given by $2q\Omega^2 x\hat{\mathbf{x}}$ in eqn. (2), provides a persistent, large-scale forcing. Eqn. (2) also has another velocity-dependent source term given by the Coriolis force, $2\mathbf{\Omega}\times\mathbf{v}$. Because the latter source term has a velocity-dependence, it contributes power on all scales. However, the smaller scales in a driven turbulence have decreasing amounts of power, with the result that we expect the second source term to provide much of its power on larger scales. Note too that the shear across the computational domain, $q\,\Omega\,H$, is larger than the flow velocities that develop in an MRI simulation with the result that the centripetal term dominates. Consequently, we would expect the velocity field in an MRI simulation to display the characteristics of a driven turbulence with large-scale forcing.

Fig. 2 shows that the toroidal magnetic field develops structures that can be comparable to the outer scales of the turbulence. This is due to the large-scale stretching in the toroidal direction by the mean shear in the flow. The poloidal magnetic field could have a dynamo-like contribution arising from its interaction with the large-scale toroidal field and that contribution could act at all length scales. Thus, we do not expect an inertial range to develop at intermediate wave numbers in the spectrum of the poloidal magnetic field. An alternative line of reasoning stems from the work of (Brandenburg 2001, Balbus & Henri 2008). It consists of saying that if Pm ≥ 1 in some of our better simulations then we would expect a buildup of small scale magnetic energy which cascades back to the larger scales. When an inertial range is established at least in the velocity for an MRI simulation with a fixed Pm, we expect $\alpha_{Tot}$ to become a scale-free parameter of the turbulence. FPLH07 develop the equivalent expectation that at large values of Re, which are essential for producing a well-developed turbulence, the critical value for the magnetic Prandtl number becomes independent of the Reynolds number. If that ansatz were true, $\alpha_{Tot}$ in our higher resolution simulations must be regulated only by Pm which is, in turn, determined by the Riemann solver in higher order Godunov schemes. Higher numerical resolution results in increasing values of the mesh Reynolds number. Our higher resolution simulations also show that $\alpha_{Tot}$ seems to become weakly dependent on resolution, i.e. of the mesh Reynolds number. Furthermore, at those higher resolutions, our $\alpha_{Tot}$ seems to depend only on the kind of Riemann solver that was used in the simulation.

To make the above facts fit into a coherent whole, all that is required is a demonstration that the velocity field develops an inertial range in our MRI-driven turbulence simulations. Figs. 3a and 3b show the power spectrum for the poloidal velocity from runs LW192 and LW128 respectively. The solid line shows the power spectrum averaged over several simulated data cubes that are evenly spaced from the fortieth orbit to the end of the simulation. The dashed lines show one standard deviation in the data at each wavenumber. While the standard deviation can become large at the outer, i.e. forcing scales, we see that it is indeed very small on all the other scales including the intermediate scales where we expect an inertial range to develop. The straight line, when it is present, shows the Kolmogorov scaling, $k^{-5/3}$. Kolmogorov scaling is well-justified in MRI simulations owing to the subsonic speeds and the modest amount of compressibility. Figs. 3c and 3d show compensated spectra for the poloidal velocity from runs LW192 and LW128 respectively, where the compensation follows Kolmogorov scaling. Figs. 3e and 3f show the power spectrum for the toroidal magnetic field from runs LW192 and LW128 respectively. Figs. 3g and 3h show the power spectrum for the poloidal magnetic field from runs LW192 and LW128 respectively. The velocity field in the $128^3$ zone simulation does not display an inertial range as can be seen from Figs. 3b and 3d. However, we do see the suggestion of a small inertial range in the velocity power spectrum from the $192^3$ zone simulation, as can be seen from Figs. 3a and 3c. The compensated spectra in Figs. 3c and 3d make it easier to deduce that there is a growing inertial range with increasing resolution. This is a most interesting result and should be confirmed, in time, with higher resolution simulations. It shows, among other things, that a flow with a high numerical Reynolds number has developed in the simulations. We do not fully understand why there is no discernable inertial range in the highest resolution velocity spectra of FP07. The lack of an inertial range might be taken to imply that a high enough numerical Reynolds number was perhaps not achieved in those simulations. It is, therefore, possible that the absence of an inertial range in FP07 might explain the resolution-dependent values of $\alpha_{\mathrm{Tot}}$ that were observed at all resolutions in that paper.

The spectra for the toroidal magnetic field in Figs. 3e and 3f do not show the formation of an inertial range. Correspondingly, the spectrum does not form a power law (at any of the resolutions used here). However, the toroidal magnetic fields are constantly re-energized by the presence of the mean shear in the simulations. Comparing the spectra for the toroidal and poloidal magnetic fields, we see that the poloidal magnetic fields have a lot less power on the largest scales, a result that is consistent with the shearing motions. It is entirely possible that the Coriolis force causes energy transfers to take place from the toroidal to the poloidal magnetic field. We will see in the next Sub-section that such a trend is indicated in our study of the transfer functions that can be obtained from our simulations.

## V.2) Quantifying Numerical Dissipation

We now wish to quantify the magnetic Reynolds number, $\text{Re}_M \equiv c_s H/\eta$, in our simulations. To do this, we first decompose the velocity into a turbulent velocity and a mean shearing velocity $\mathbf{v} = \mathbf{v}_t + \mathbf{V}_{sh}$. Here $\mathbf{V}_{sh} \equiv -2q\Omega x \hat{\mathbf{y}}$ carries the mean shear. We now rewrite eqn. (3) as

$$\frac{d\mathbf{B}}{dt} = -V_{sh}\frac{\partial \mathbf{B}}{\partial y} + B_x \frac{\partial V_{sh}}{\partial x}\hat{\mathbf{y}} - (\mathbf{v}_t \cdot \nabla)\mathbf{B} - (\nabla \cdot \mathbf{v}_t)\mathbf{B} + (\mathbf{B} \cdot \nabla)\mathbf{v}_t. \tag{8}$$

Following FP07, we take the Fourier transform of eqn. (8) and multiply by the complex conjugate of the magnetic field $\tilde{\mathbf{B}}^*$, we can recast our equation in spectral space as

$$\frac{1}{2}\frac{\partial |\mathbf{B}(\mathbf{k})|^2}{\partial t} = A + S + T_{bb} + T_{divv} + T_{bv}, \tag{9}$$

where

$$A = -\Re\left[\tilde{\mathbf{B}}^*(\mathbf{k}) \cdot \iiint V_{sh}\frac{\partial \mathbf{B}}{\partial y}e^{-i\mathbf{k}\cdot\mathbf{x}}d^3\mathbf{x}\right], \tag{10}$$

$$S = \Re\left[\tilde{B}_y^*(\mathbf{k}) \cdot \iiint B_x\frac{\partial V_{sh}}{\partial x}e^{-i\mathbf{k}\cdot\mathbf{x}}d^3\mathbf{x}\right], \tag{11}$$

$$T_{bb} = -\Re\left[\tilde{\mathbf{B}}^*(\mathbf{k}) \cdot \iiint [(\mathbf{v}_t \cdot \nabla)\mathbf{B}]e^{-i\mathbf{k}\cdot\mathbf{x}}d^3\mathbf{x}\right], \tag{12}$$

$$T_{divv} = -\Re\left[\tilde{\mathbf{B}}^*(\mathbf{k}) \cdot \iiint [(\nabla \cdot \mathbf{v}_t)\mathbf{B}]e^{-i\mathbf{k}\cdot\mathbf{x}}d^3\mathbf{x}\right], \tag{13}$$

$$T_{bv} = \Re\left[\tilde{\mathbf{B}}^*(\mathbf{k}) \cdot \iiint [(\mathbf{B} \cdot \nabla)\mathbf{v}_t]e^{-i\mathbf{k}\cdot\mathbf{x}}d^3\mathbf{x}\right]. \tag{14}$$

In computing the values of eqns. (10) to (14), we have followed the remap procedure of Hawley et al. (1995) to account for the shear. Here $\Re$ denotes the real parts of the spectral quantities in the above equations. The shear will cause the length scale to be larger in the $y$ direction than in the $x$ and $z$ directions, so we will only consider the $k_y = 0$ plane in spectral space. The terms in eqns. (10) to (14) are known as transfer functions and indicate the direction in which energy flows in a fully developed, steady-state turbulence. The term $A$ represents the role of the stretched toroidal field. Because the net flux is zero and we consider only the $k_y = 0$ plane, we have that $A = 0$. $S$ represents the role of shear and is zero when only poloidal fields are considered. The transfer function $T_{bb}$ characterizes the transfer of magnetic energy to smaller scales. The transfer function $T_{\text{div}v}$ catalogues the role of compressibility and $T_{bv}$ describes the role of field line stretching due to the turbulent flow field.

In steady state we expect the magnetic energy to saturate, with the result that $\partial |\mathbf{B}(\mathbf{k})|^2 / \partial t$ in eqn. (9) will be zero. The right hand side of eqn. (9) is also expected to be zero in steady state. Notice, though, that eqn. (9) only represents the situation as it would prevail in an idealized, analytic calculation. Any numerical code will have some numerical dissipation that is required to stabilize the smallest scale modes. Without such stabilization, the numerical code would become unstable as soon as discontinuities of any sort develop in a calculation. Restricting our attention to the poloidal magnetic field in steady state, we therefore expect

$$T_{bb}^{\text{P}} + T_{\text{div}v}^{\text{P}} + T_{bv}^{\text{P}} + D_{num}^{\text{P}} = 0 \tag{15}$$

where $D_{num}^{\text{P}}$ is the numerical dissipation provided by the code and the superscript "P" denotes that only the poloidal part of the magnetic field is being considered. By accounting for the first three terms in the poloidal part of the flow, we are afforded a mechanism for teasing out the spectral dependence of $D_{num}^{\text{P}}$ from eqn. (15). On the smallest scales, we expect the dissipation that the Riemann solver provides to mimic a physical resistivity. We denote this physical dissipation by $D_{res}(k)$ which is defined by

$$D_{res}(k) \equiv \eta\, k^2 \left| \tilde{\mathbf{B}}^P(k) \right|^2 \qquad (16)$$

where $\tilde{\mathbf{B}}^P(k)$ is the spectral representation of the poloidal magnetic field.

Fig. 4a shows $T_{bb}^P$ (dashed line), $T_{divv}^P$ (dotted line) and $T_{bv}^P$ (dot-dash line) as a function of wave number for LW192. A solid horizontal line with an ordinate of zero is provided as a point of reference in Fig. 4a. Fig. 4b plots the resulting $D_{num}^P$ as a function of wave number (solid line) for LW192. It also plots $D_{res}(k)$ as a dashed line. A dotted horizontal line with an ordinate of zero is provided as a point of reference in Fig. 4b. Figs. 4c and 4d do the same for LW128, Figs. 4e and 4f do the same for LW96 and Figs 4g and 4h do the same for LW64. The spectra in Fig. 4 represent time-averages from simulation dumps that were evenly spaced from the fortieth orbit to the end of the simulation. Fig. 4a shows us that $T_{bv}^P$ is positive for a large range of length scales showing that field line stretching operates over a range of length scales in the simulation due to forcing provided by the MRI. $T_{bb}^P$ is also positive on the largest scales, indicating that the large scale toroidal field transfers part of its energy to the poloidal field via a dynamo-like action. $T_{divv}^P$ is negative through most of the spectral range, indicating that the compressibility in the flow only contributes a forward cascade of magnetic energy. Similar trends are observed in Figs. 4c, 4e and 4g though all of the trends are not as pronounced with decreasing resolution.

The solid line in Fig. 4b shows the variation of $D_{num}^P$ as a function of wave number. The dashed line in Fig. 4b represents the best fit we could obtain at large wave numbers between $D_{res}(k)$ and $D_{num}^P$. This fit was obtained by tuning the value of $\eta$ in eqn. (16) so as to obtain a good match up between $D_{res}(k)$ and $D_{num}^P$ at large wave numbers. Such a fitting procedure allows us to obtain a time-averaged value for the *effective, time-averaged* resistivity that was invoked by the code in the particular simulation. Please note though that this is not the actual resistivity of the higher order

Godunov scheme. Table III catalogues the effective values of $Re_M$ obtained from such a fitting procedure applied to all our LW and LC simulations in Fig. 4. Denoting the maximum wave number in a simulation by $k_{max}$, the fitting procedure consisted of a least squares minimization of the difference between $D_{res}(k)$ and $D_{num}^P$ in the range $[k_{max}/2, k_{max}]$. We see from Table III that the simulations develop effective magnetic Reynolds numbers of the order of several tens of thousands. Furthermore, the effective magnetic Reynolds number increases with linear resolution showing that as the resolution is increased the overall dissipation decreases. Fig. 5 plots the values of $Re_M$ from Table III as a function of the number of mesh points in any one direction in our simulations. Both the LW and LC runs show a linear variation of the magnetic Reynolds number with increasing number of zones with the LW runs showing a somewhat higher slope than the LC runs. Our present method of obtaining a magnetic Reynolds number may be thought of as imperfect, however, the problem is strongly non-linear and the approximate demonstration provided here is quite suggestive and consistent with expected trends.

Fig. 4b shows that the dissipation in our LW192 simulation acts on a very small and concentrated range of high wave numbers. This is exactly what one desires from a high resolution shock capturing scheme. Such schemes are optimal in restricting the numerical dissipation to very small-scale structures while attempting to leave the larger scales relatively free of dissipation. Despite the larger resolution of the simulation reported in Fig. 9 of FP07, we see that their numerical dissipation operates on a much larger range of scales. Unlike the long wavelength oscillations in Fig. 9 of FP07, our $D_{num}^P$ is uniformly positive on larger scales in Fig. 4b, indicating that the large scale forcing terms provided by the transfer functions $T_{bb}^P$ and $T_{bv}^P$ are driving an inverse cascade of magnetic energy on those scales.

FP07 obtained a nominal value for their $Re_M$ that is larger than ours despite the fact that their code was dissipative over a much larger range of wave numbers. It is interesting to ask how this could come about. We point out that if a scheme does not dissipate the energy in a turbulent cascade on larger scales, then that same energy cascades to the smaller scales and has to be dissipated even more strongly on those scales. In reality, for these self-adjusting schemes, the effective value of $\eta$ for any wave number $k$ depends on the wave number as well as the amplitude of the fluctuations at that wave number. Furthermore, $D_{res}(k)$ varies in response to this wave-number dependent $\eta$. Eqn. (16) makes a considerable simplification by assuming a constant $\eta$. Any procedure that attempts to match $D_{res}(k)$ to $D_{num}^{P}$ at large wave numbers will necessarily produce a larger value of $\eta$ for a scheme that restricts its dissipation to smaller scales. Thus the procedure for obtaining $Re_M$ that was presented in FP07, and carried over to this paper, is useful when comparing a single numerical scheme at different resolutions. However, it is not very useful when inter-comparing different numerical schemes. In such situations, a direct spectral plot of the numerical dissipation $D_{num}^{P}$ proves more useful.

It is interesting to cross-compare different reconstruction strategies when the same Riemann solver is used. Figs. 6a and 6b plot the dissipation $D_{num}^{P}$ (solid line) as well as $D_{res}(k)$ (dashed line) a function of wave number for LC192 and LM192 respectively. They should be compared to Fig. 4b. We see that LW and LC have rather similar dissipation characteristics, with the LW showing a slightly larger range of positive values that LC. This indicates that the WENO reconstruction has permitted the growing modes of the MRI to operate undamped over a larger range of intermediate and large wavelengths compared to the MC reconstruction. The numerical dissipation in the LW192 run is confined to a very small range of short wavelengths. These trends should be contrasted with the plot for the LM192 calculation which shows that only a small range of long wavelength modes are allowed to grow by the MinMod reconstruction, while the remaining modes are strongly damped over a large range of wavelengths. Fig. 6b should be compared to Fig. 9 from FP07. It shows that the dissipation characteristics of the numerical code used in FP07, at least as they pertain to this problem, are roughly similar to those of a Godunov scheme that uses a very modest reconstruction strategy.

The schemes used here are self-adjusting in that they examine the flow structures and automatically adjust the dissipation so as to avoid spurious oscillations from forming. If the flow is smooth, the Godunov schemes produce as little numerical dissipation as is needed for numerical stability. If the flow has strong discontinuities, a Godunov code can supply as much additional dissipation as is needed to prevent any further spurious oscillations at the location of the discontinuity. It is, therefore, not possible to simulate a simpler problem, like the propagation of Alfven waves, and use it to deduce the effective viscosities and resistivities that develop in the code (Lesaffre & Balbus 2007). Evaluating transfer functions for the terms in the momentum equation may provide one way of calculating the *effective, time-averaged* viscosities that develop in our MRI simulations. Since such methods for obtaining the effective, time-averaged numerical viscosity have not been developed for MRI simulations, an evaluation of the viscosity by evaluating transfer functions is out of the scope of this work. However, one can in general say that higher order Godunov codes, as well as the Riemann solvers they draw on, treat velocities and magnetic fields on an equal footing. As a result, the magnetic Prandtl numbers will usually be of the order of unity. This allows us to justify our claim in Section III that the effective, numerical Reynolds numbers achieved by such simulations tend to be rather large.

**VI) Discussion and Conclusions**

Higher order Godunov codes usually draw on a variety of algorithmic elements in order to obtain a high quality result. Thus, one can draw on several different types of reconstruction strategies. One can also use a range of Riemann solvers. These two algorithmic pieces constitute the most important building blocks for schemes of this type. Different time-stepping strategies that one might use in second order Godunov schemes also constitute an important building block. However, a large body of prior experience tells us that differences in time-stepping strategies do not produce much of a difference in the outcomes for second order accurate Godunov schemes. In this paper we have studied the role played by different reconstruction strategies and different Riemann solvers in simulations of MRI-driven turbulence. We find that details of the Riemann solvers change the effective Prandtl number in the calculation and, therefore, strongly influence outcomes. The different reconstruction strategies do not have such a pronounced influence on the outcomes. However, higher quality reconstruction strategies are shown to be very important in avoiding channel modes and producing a vigorous MRI-driven turbulence on smaller meshes. We conclude that the better reconstruction strategies (MC or WENO) and the better Riemann solvers (HLLD or linearized) produce vigorous MRI-driven turbulence.

The dependence of the effective $\alpha$ on numerical resolution is also explored. We show that the effective $\alpha$ parameter tends to show progressively smaller decrements with increasing resolution when the best reconstruction strategy (WENO) and the best Riemann solver (linearized) are used. Spectra for the poloidal velocity field show the formation of a very small inertial range in the largest simulations presented here. The presence of an inertial range suggests that a high effective Reynolds number develops in our simulations. The magnetic spectra do not seem to form an inertial range at any of the resolutions tested here and this is consistent with the inverse cascade that takes place in a calculation with magnetic Prandtl number exceeding unity. The toroidal magnetic field is shown to have more energy on larger scales than the poloidal magnetic field, a result that is consistent with the persistent shear in the toroidal direction in MRI simulations. There is a significant transfer of magnetic energy to larger scales taking place over a range of scales in these simulations; a conclusion that is also supported by our study of transfer functions.

Transfer functions have been evaluated for the poloidal magnetic field. Taken in conjunction with the fact that a steady-state turbulence has been established, the transfer functions enable us to calculate an effective, time-averaged magnetic Reynolds number for our simulations. The effective magnetic Reynolds numbers are found to be quite large and they are shown to increase with increasing linear resolution. This is the expected behavior from a well-resolved simulation. A similar effort to derive transfer functions for the velocity fields would yield rich dividends in that it would enable us to deduce the effective, time-averaged Reynolds number for our simulations. As it stands, we can rely on the fact that the magnetic Prandtl numbers in these simulations are usually of the order of unity, to deduce that the effective, time-averaged Reynolds number in our larger simulations should also be quite large. The emergence of an inertial range in the poloidal velocity provides further support to our claim that large effective, time-averaged Reynolds numbers develop in some of our larger simulations.

At large enough Reynolds numbers our simulations, along with the work of FPLH07, give support to the hypothesis that the critical magnetic Prandtl number might become independent of the Reynolds number. The fact that our effective $\alpha$ seems to become quite independent of resolution on large enough computational meshes provides a further confirmation of that ansatz.

## Acknowledgements

The authors thank David Tilley for useful discussions. DSB acknowledges support via NSF grants AST-0607731 and NSF-AST-0947765. DSB also acknowledges NASA grants NASA-NNX07AG93G and NASA-NNX08AG69G. The majority of simulations were performed on computers run by the Center for Research Computing at UND but a few initial simulations were also performed at NASA-AMES.

**Table I.** Algorithms and mesh resolutions for each of the runs described in this paper.

| Run Name | Riemann Solver | Interpolation Scheme | Resolution |
|---|---|---|---|
| LW192 | Linearized | WENO | $192^3$ |
| LW128 | | | $128^3$ |
| LW96 | | | $96^3$ |
| LW64 | | | $64^3$ |
| LC192 | | MC | $192^3$ |
| LC128 | | | $128^3$ |
| LC96 | | | $96^3$ |
| LC64 | | | $64^3$ |
| LM192 | | MinMod | $192^3$ |
| LM128 | | | $128^3$ |
| LM96 | | | $96^3$ |
| LM64 | | | $64^3$ |
| DW128 | HLLD | WENO | $128^3$ |
| DW96 | | | $96^3$ |
| DW64 | | | $64^3$ |
| DC128 | | MC | $128^3$ |
| DC96 | | | $96^3$ |
| DC64 | | | $64^3$ |
| DM128 | | MinMod | $128^3$ |
| DM96 | | | $96^3$ |
| DM64 | | | $64^3$ |
| EW128 | HLLE | WENO | $128^3$ |
| EW64 | | | $64^3$ |
| EM128 | | MinMod | $128^3$ |
| EM64 | | | $64^3$ |

**Table II.** Summary of scalar values averaged from orbit 40 to the end of the simulation. All values are $\times 10^{-3}$. A "C" in the $\alpha_{\text{Tot}}$ column indicates that this run has developed a channel mode.

| Run | $\frac{1}{8\pi P_0}\langle B^2 \rangle$ | $\frac{1}{P_0}\left\langle \frac{1}{2}\rho\left(v_x^2 + v_z^2\right) \right\rangle$ | $\alpha_{\text{Max}}$ | $\alpha_{\text{Rey}}$ | $\alpha_{\text{Tot}}$ |
|---|---|---|---|---|---|
| LW192 | 6.4 ± 1.2 | 1.5 ± 0.3 | 2.8 ± 0.5 | 0.6 ± 0.3 | 3.4 ± 0.7 |
| LW128 | 6.9 ± 1.0 | 1.7 ± 0.3 | 3.1 ± 0.5 | 0.7 ± 0.3 | 3.8 ± 0.6 |
| LW96 | 9.7 ± 1.7 | 2.6 ± 0.5 | 4.2 ± 0.7 | 1.0 ± 0.3 | 5.3 ± 0.9 |
| LW64 | 9.9 ± 1.8 | 2.6 ± 0.5 | 4.2 ± 0.7 | 1.0 ± 0.4 | 5.2 ± 1.0 |
| LC192 | 5.7 ± 1.0 | 1.5 ± 0.3 | 2.5 ± 0.4 | 0.6 ± 0.2 | 3.1 ± 0.6 |
| LC128 | 7.1 ± 0.8 | 1.9 ± 0.2 | 3.2 ± 0.3 | 0.7 ± 0.2 | 3.9 ± 0.5 |
| LC96 | 8.7 ± 1.8 | 2.4 ± 0.5 | 3.9 ± 0.8 | 0.9 ± 0.4 | 4.8 ± 1.0 |
| LC64 | 9.9 ± 1.6 | 2.9 ± 0.7 | 4.3 ± 0.7 | 1.1 ± 0.4 | 5.4 ± 1.0 |
| LM192 | 6.4 ± 1.0 | 1.4 ± 0.2 | 2.8 ± 0.4 | 0.6 ± 0.2 | 3.5 ± 0.6 |
| LM128 | 7.1 ± 1.5 | 1.5 ± 0.4 | 3.0 ± 0.7 | 0.7 ± 0.2 | 3.7 ± 0.9 |
| LM96 | 3.8 ± 1.0 | 0.7 ± 0.2 | 1.4 ± 0.4 | 0.3 ± 0.1 | 1.7 ± 0.6 |
| LM64 | 0.6 ± 0.4 | 2.2 ± 2.9 | 0.1 ± 0.1 | 0.0 ± 1.4 | 0.0 ± 1.4c |
| DW128 | 8.0 ± 1.4 | 2.0 ± 0.4 | 3.5 ± 0.6 | 0.8 ± 0.2 | 4.4 ± 0.7 |
| DW96 | 11.0 ± 2.0 | 2.6 ± 0.5 | 4.8 ± 0.9 | 1.1 ± 0.4 | 5.9 ± 1.1 |
| DW64 | 11.0 ± 2.0 | 2.9 ± 0.5 | 4.6 ± 0.8 | 1.2 ± 0.4 | 5.8 ± 1.1 |
| DC128 | 8.0 ± 1.2 | 2.1 ± 0.3 | 3.6 ± 0.5 | 0.8 ± 0.3 | 4.4 ± 0.7 |
| DC96 | 8.8 ± 1.2 | 2.5 ± 0.4 | 4.0 ± 0.6 | 1.0 ± 0.3 | 4.9 ± 0.7 |
| DC64 | 12.4 ± 2.5 | 3.5 ± 0.9 | 5.3 ± 1.1 | 1.3 ± 0.6 | 6.6 ± 1.4 |
| DM128 | 8.3 ± 1.6 | 1.7 ± 0.4 | 3.5 ± 0.7 | 0.8 ± 0.2 | 4.3 ± 0.8 |
| DM96 | 6.0 ± 0.9 | 1.2 ± 0.2 | 2.4 ± 0.4 | 0.6 ± 0.1 | 2.9 ± 0.5 |
| DM64 | 0.6 ± 0.4 | 1.0 ± 1.4 | 0.1 ± 0.1 | 0.0 ± 0.9 | 0.1 ± 0.9c |
| EW128 | 1.8 ± 1.3 | 0.8 ± 0.5 | 0.6 ± 0.6 | 0.1 ± 2.3 | 0.7 ± 2.4c |
| EW64 | 0.3 ± 0.4 | 3.5 ± 2.6 | 0.0 ± 0.0 | 0.1 ± 3.5 | 0.1 ± 3.5c |
| EM128 | 4.0 ± 1.0 | 0.1 ± 0.1 | 0.0 ± 0.0 | 0.0 ± 0.9 | 0.0 ± 0.9c |
| EM64 | 0.0 ± 0.0 | 0.2 ± 0.3 | 0.0 ± 0.0 | 0.0 ± 0.2 | 0.0 ± 0.2c |

**Table III** catalogues the values of $\mathrm{Re}_M$ obtained by using the fitting procedure described in Section V.2 which is applied to all our LW and LC simulations in Fig. 4.

| Run Name | $\mathrm{Re}_M$ |
|---|---|
| LW192 | $3.40 \times 10^4$ |
| LW128 | $2.05 \times 10^4$ |
| LW96 | $1.17 \times 10^4$ |
| LW64 | $7.10 \times 10^3$ |
| LC192 | $2.95 \times 10^4$ |
| LC128 | $1.90 \times 10^4$ |
| LC96 | $1.20 \times 10^4$ |

| LC64 | $5.50 \times 10^3$ |

**Figures**

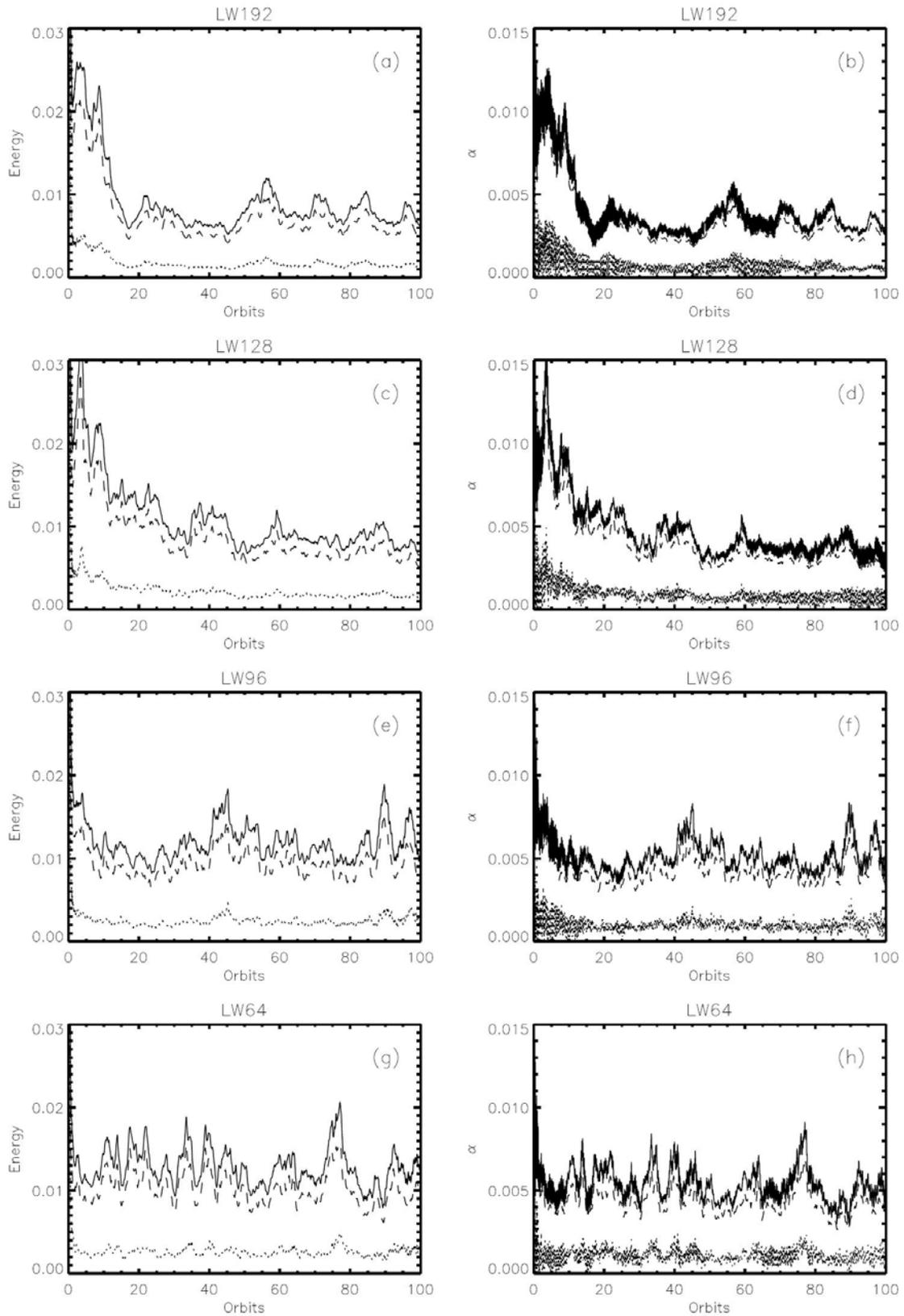

Figs. 1a, 1c, 1e and 1g show the variation of the kinetic energy (dotted line), magnetic energy (dashed line) and total energy (solid line) as a function of time in the LW192, LW128, LW96 and LW64 simulations respectively. Figs. 1b, 1d, 1f and 1h show the variation of $\alpha_{Rey}$ (dotted line), $\alpha_{Max}$ (dashed line) and $\alpha_{Tot}$ (solid line) as a function of time in the LW192, LW128, LW96 and LW64 simulations respectively.

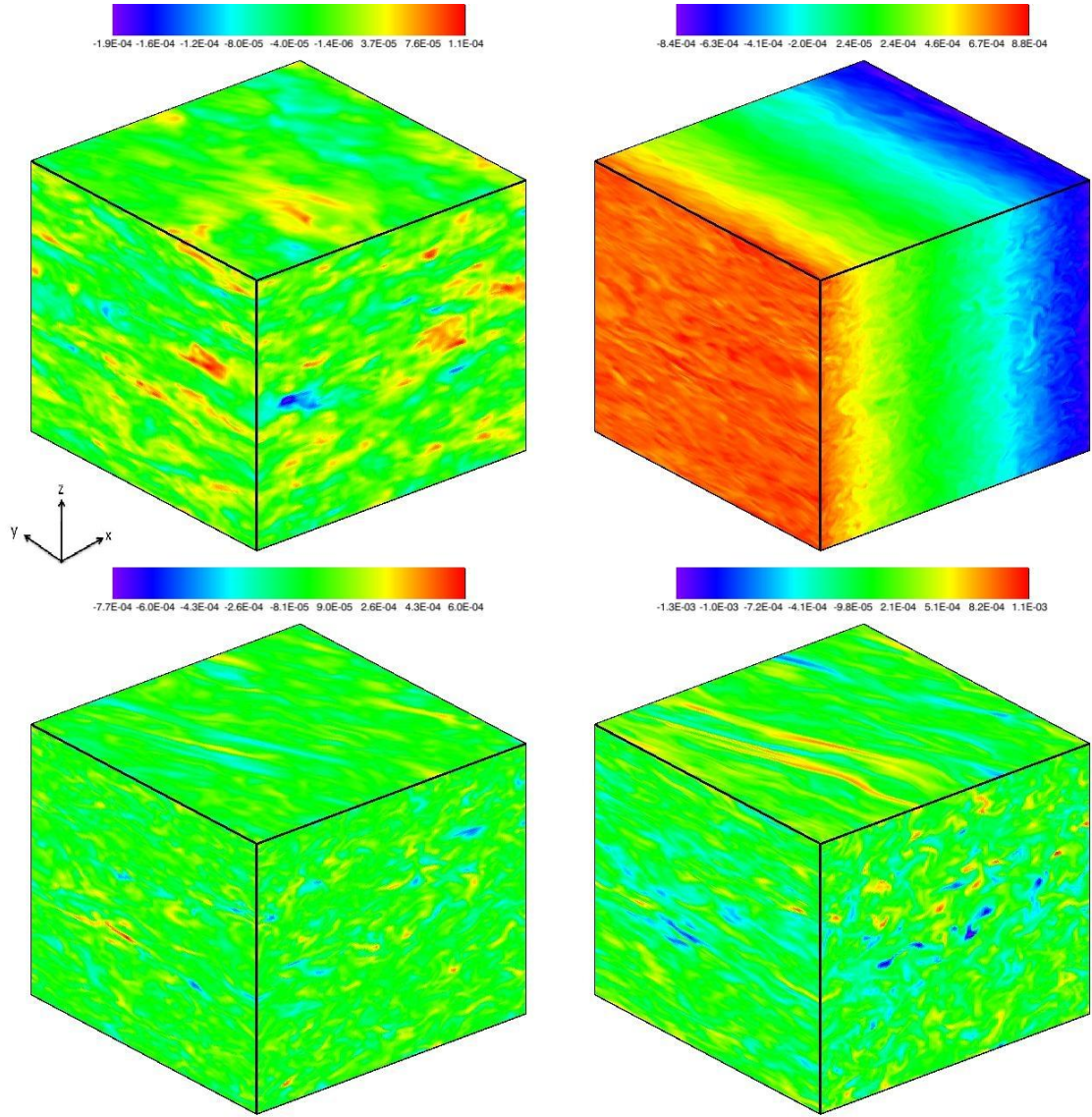

Fig. 2. From left to right, top to bottom, the figures show a snapshot of the magnitudes of $\mathbf{v}_x$, $\mathbf{v}_y$, $\mathbf{B}_x$ and $\mathbf{B}_y$ respectively for LW192 after eighty orbits have been simulated. The x, y and z-axes show the radial, azimuthal and vertical directions in the accretion disk. The flow variables on the three faces of the computational domain that are nearest to the reader are shown. Notice the active turbulence. The mean shearing flow is clearly seen in Fig. 2b. In each of the snapshots, the left panel represents the $x = -H/2$ plane, the right panel is the $y = H/2$ plane, and the top panel is the $z = H/2$ plane.

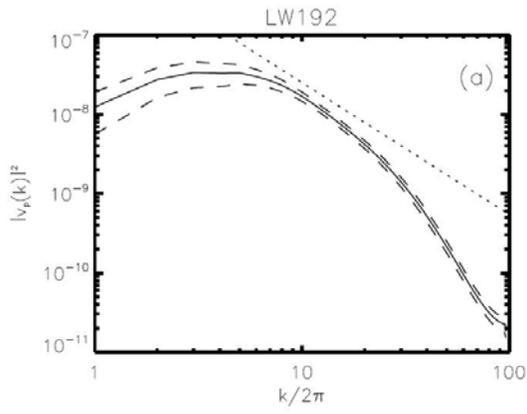
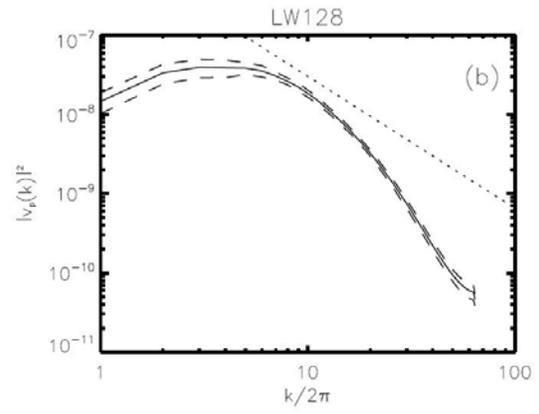
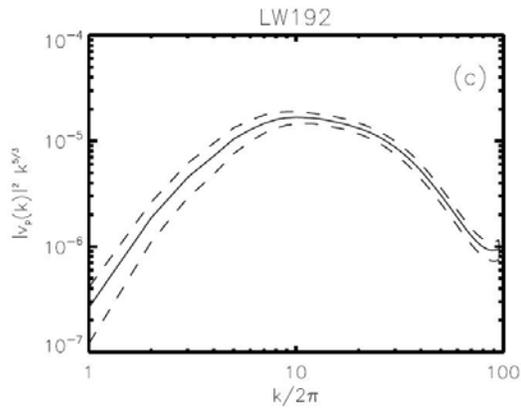
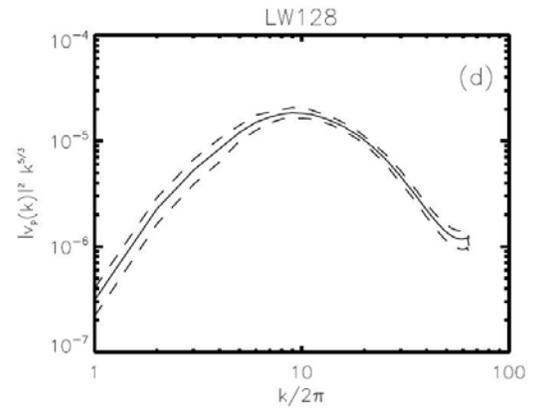
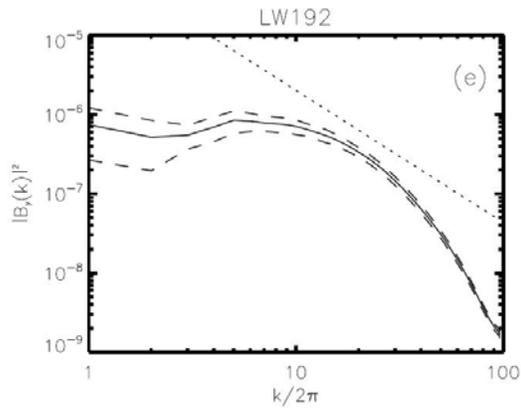
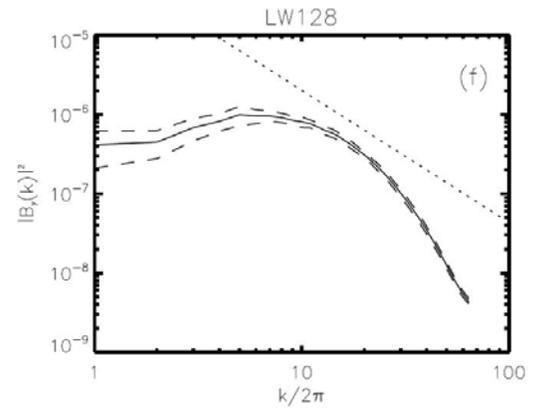
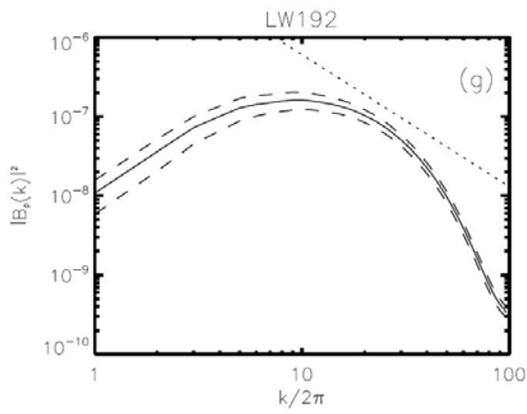
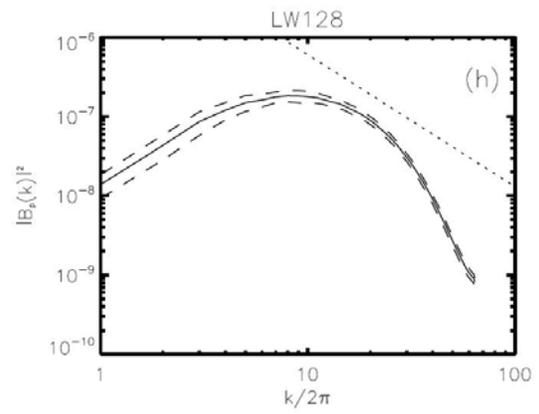

Figs. 3a and 3b show the power spectrum for the poloidal velocity from runs LW192 and LW128 respectively. The solid line shows the power spectrum averaged over several simulated data cubes that are evenly spaced from the fortieth orbit to the end of the simulation. The dashed lines show one standard deviation in the data at each wavenumber. The straight line, when it is present, shows the Kolmogorov scaling, $k^{-5/3}$. Figs. 3c and 3d show compensated spectra for the poloidal velocity from runs LW192 and LW128 respectively, where the compensation follows Kolmogorov scaling. Figs. 3e and 3f show the power spectrum for the toroidal magnetic field from runs LW192 and LW128 respectively. Figs. 3g and 3h show the power spectrum for the poloidal magnetic field from runs LW192 and LW128 respectively.

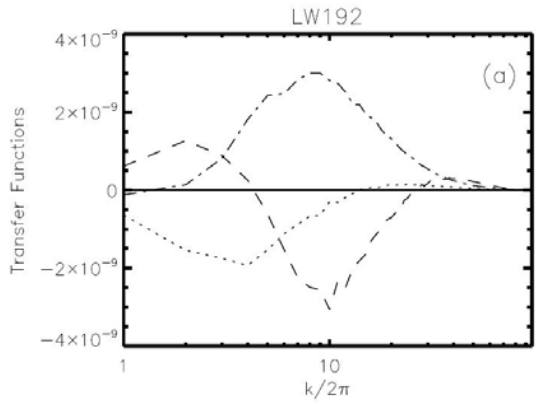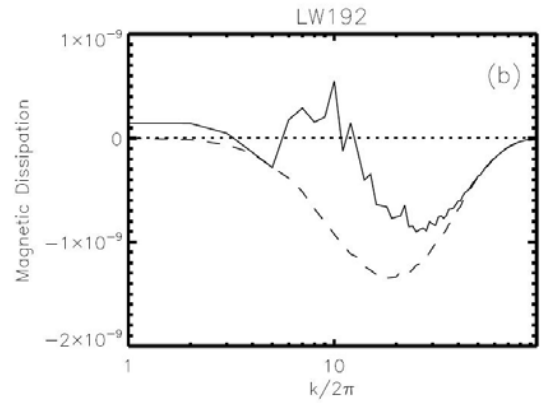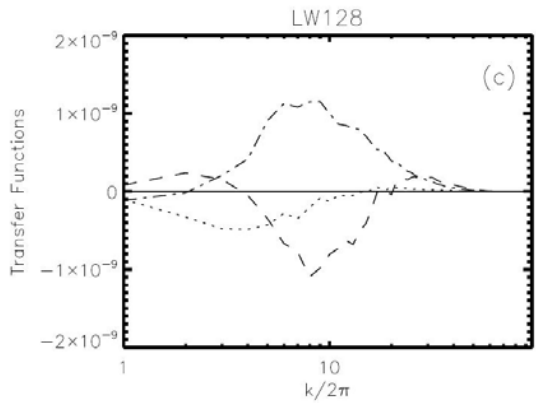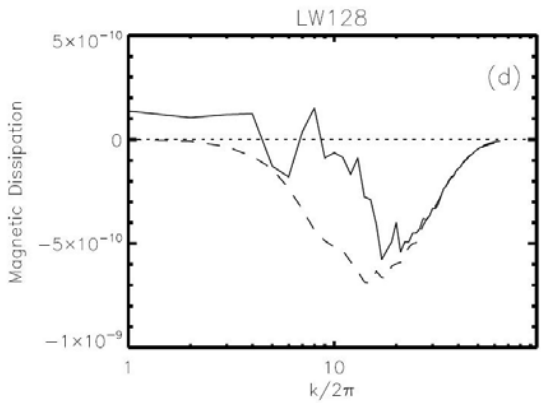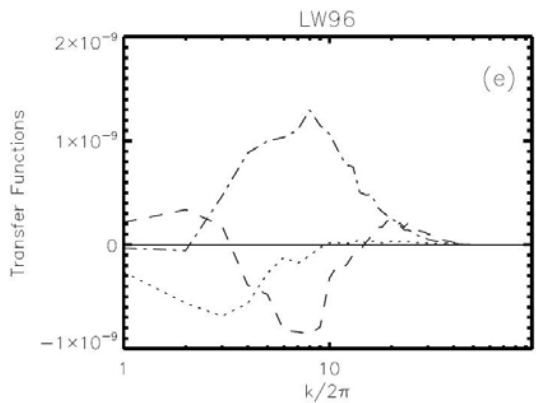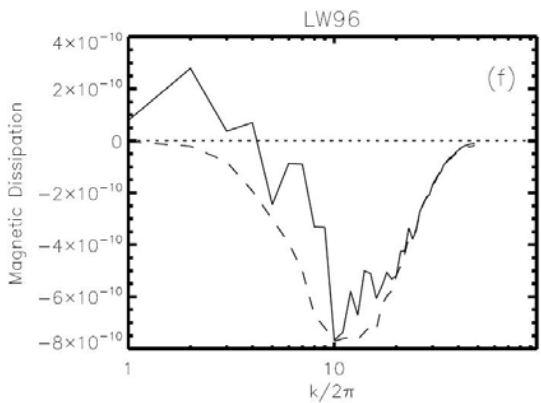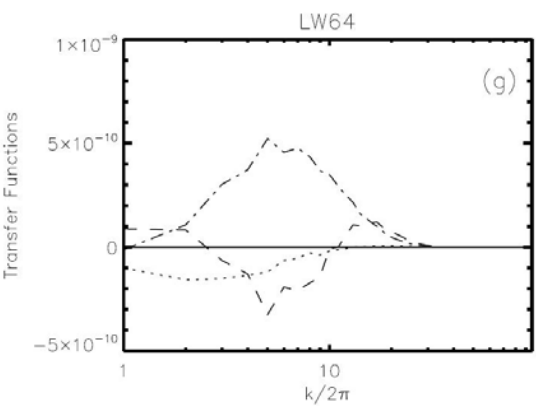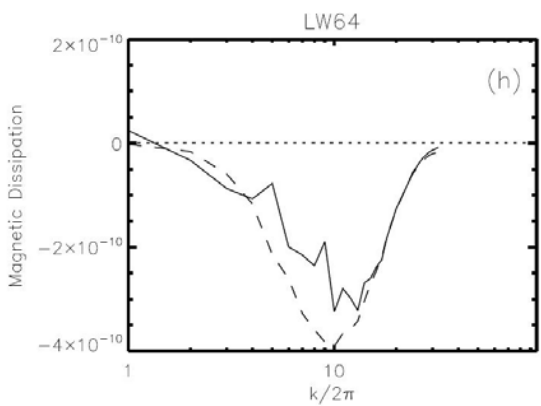

Fig. 4a plots $T^\text{P}_{bb}$ (dashed line), $T^\text{P}_\text{divv}$ (dotted line) and $T^\text{P}_{bv}$ (dot-dash line) as a function of wave number for LW192. A solid horizontal line with an ordinate of zero is provided as a point of reference in Fig. 4a. Fig. 4b plots the resulting $D_\text{num}$ as a function of wave number (solid line) for LW192. It also plots $D_\text{res}(k)$ as a dashed line. A dotted horizontal line with an ordinate of zero is provided as a point of reference in Fig. 4b. Figs. 4c and 4d do the same for LW128, Figs. 4e and 4f do the same for LW96 and Figs 4g and 4h do the same for LW64.

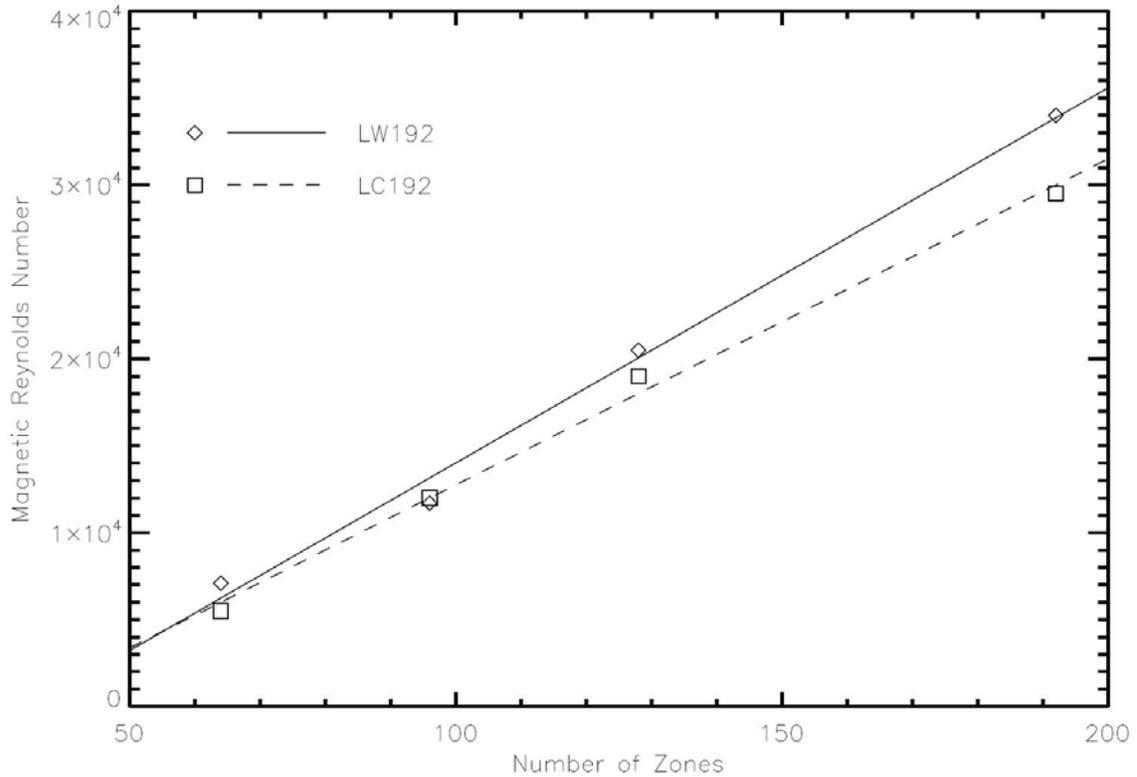

Fig. 5 plots the values of Re$_\text{M}$ as a function of the number of mesh points in any one direction in our simulations.

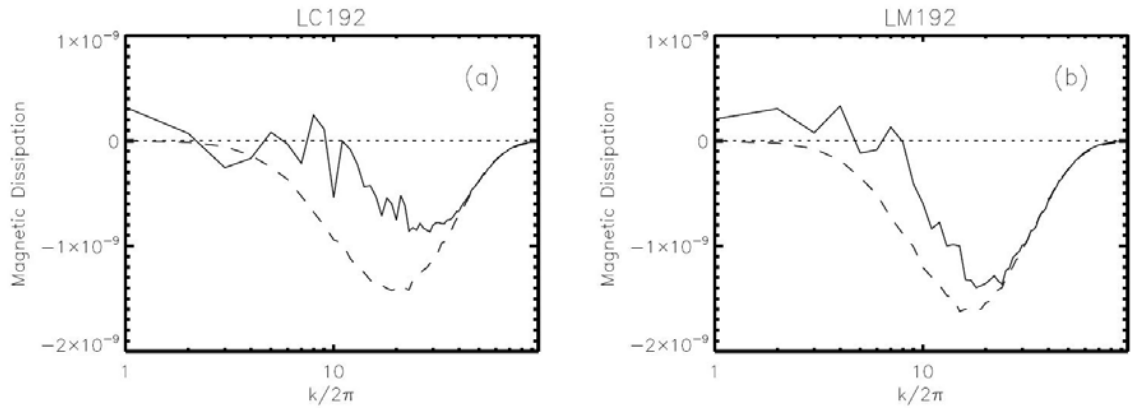

Figs. 6a and 6b plot the dissipation $D_{num}^{P}$ (solid line) as well as $D_{res}(k)$ (dashed line) a function of wave number for LC192 and LM192 respectively.